\documentclass{ieeetj}
\usepackage{cite}
\usepackage{amsmath,amssymb,amsfonts}
\usepackage{algorithmic}
\usepackage{graphicx,color}
\usepackage{textcomp}
\usepackage{xcolor}
\usepackage{subfig}
\usepackage{hyperref}
\usepackage{threeparttable}
\usepackage{makecell}
\usepackage{stfloats}
\usepackage{amssymb}    
\usepackage{bbding}     
\usepackage{pifont}
\usepackage{booktabs}
\usepackage{array}
\usepackage{siunitx}
\usepackage{xspace}
\usepackage{ragged2e}
\usepackage{amssymb}
\usepackage{ragged2e}
\hypersetup{hidelinks=true}
\usepackage{algorithm,algorithmic}
\def\BibTeX{{\rm B\kern-.05em{\sc i\kern-.025em b}\kern-.08em
    T\kern-.1667em\lower.7ex\hbox{E}\kern-.125emX}}
\AtBeginDocument{\definecolor{tmlcncolor}{cmyk}{0.93,0.59,0.15,0.02}\definecolor{NavyBlue}{RGB}{0,86,125}}

\def\OJlogo{\vspace{-4pt}$<$Society logo(s) and publication title will appear here.$>$}
\def\seclogo{\vspace{10pt}$<$Society logo(s) and publication title will appear here.$>$}

\def\authorrefmark#1{\ensuremath{^{\textbf{#1}}}}

\begin{document}
\receiveddate{XX Month, XXXX}
\reviseddate{XX Month, XXXX}
\accepteddate{XX Month, XXXX}
\publisheddate{XX Month, XXXX}
\currentdate{XX Month, XXXX}
\doiinfo{XXXX.2022.1234567}

\markboth{}{R. He, M. Yang, Z. Zhang, {et al.} }

\title{Artificial Intelligence Empowered Channel Prediction: A New Paradigm for Propagation Channel Modeling}

\author{Ruisi He\authorrefmark{1}, Senior Member, IEEE, Mi Yang\authorrefmark{1}, Member, IEEE, Zhengyu Zhang\authorrefmark{1}, Student Member, IEEE, Bo Ai\authorrefmark{1}, Fellow, IEEE, and Zhangdui Zhong\authorrefmark{1}, Fellow, IEEE}
\affil{Beijing Jiaotong University, Beijing, 100044 China}
\corresp{Corresponding author: M. Yang (email: myang@bjtu.edu.cn).}

\begin{abstract}
This paper proposes a novel paradigm centered on Artificial Intelligence (AI)-empowered propagation channel prediction to address the limitations of traditional channel modeling. We present a comprehensive framework that deeply integrates heterogeneous environmental data and physical propagation knowledge into AI models for site-specific channel prediction, which referred to as channel inference. By leveraging AI to infer site-specific wireless channel states, the proposed paradigm enables accurate prediction of channel characteristics at both link and area levels, capturing spatio-temporal evolution of radio propagation. Some novel strategies to realize the paradigm are introduced and discussed, including AI-native and AI-hybrid inference approaches. This paper also investigates how to enhance model generalization through transfer learning and improve interpretability via explainable AI techniques. Our approach demonstrates significant practical efficacy, achieving an average path loss prediction root mean square error (RMSE) of $\sim$ 4 dB and reducing training time by 60\%-75\%. This new modeling paradigm provides a foundational pathway toward high-fidelity, generalizable, and physically consistent propagation channel prediction for future communication networks.
\end{abstract}


\IEEEspecialpapernotice{(Invited Paper)}

\maketitle

\section{Introduction}
\IEEEPARstart{T}{he} evolution of wireless communication systems is advancing towards the sixth generation (6G). Its core paradigm moves beyond traditional communication, aiming for deep convergence of communication, sensing, computation, and artificial intelligence (AI) to create a native-intelligent network\cite{9390169}. This vision drives expansion of application scenarios from terrestrial human-centric coverage to a holistic, three-dimensional domain encompassing seamless integration of space, air, terrestrial, and maritime environments\cite{add0108-7}. 

Wireless channel, as the foundational physical medium for signal propagation, fundamentally determines ultimate limits of communication performance, including capacity, reliability, and latency. The transition to 6G, marked by the use of mmWave, ultra-massive multiple-in multipleout (MIMO), integrated sensing and communication, and complex deployments (e.g., low-altitude drones, satellite networks), results in new propagation characteristics\cite{9267779,10500503,10982381}. This evolution in propagation channel is intrinsically and strongly coupled with the design and efficacy of communication algorithms\cite{add0108-4}. Consequently, channel modeling must evolve beyond traditional statistical or deterministic approaches. The new requirements demand a paradigm shift towards more accurate, intelligent, and efficient models with fairly high sensitivity to environment changes\cite{9237116,add0108-2}.

Wireless channel is, in essence, the product of interaction between electromagnetic waves and physical environment. Its fundamental characteristics are determined by geometry of environment, material properties, and dynamics of propagation paths\cite{10507178}. The complex propagation mechanisms results that the channel impulse response is a complex function of its surrounding environment\cite{add0108-1}. To capture such dependency, the existing channel models utilize physical environmental information in three primary approaches: i) Deterministic methods, such as ray tracing, require precise geometric and material information of environment to solve electromagnetic equations for channel prediction\cite{10419169}, ii) Stochastic methods rely on statistical distributions of channel parameters for broadly scenarios\cite{821698}, iii) Semi-deterministic methods attempt a compromise by incorporating geometric assumptions into a statistical framework\cite{10978050}. These models are simple and efficient, but fail to capture the unique environmental features and spatial characteristics of any given location, limiting the accuracy and generalizability.

To solve the above limitations, this paper proposes a novel modeling paradigm centered on AI-empowered environment feature mining and propagation channel prediction. The core idea is to use AI to fully extract heterogeneous environmental features and, by integrating priori propagation knowledge, construct an accurate and efficient mapping between multi-dimensional environmental information to propagation channel. This paradigm aims to fundamentally improve model generalization, accuracy, and computational efficiency. By extracting underlying relationships between complex environment and signal propagation, the AI-empowered approach seeks to achieve high-fidelity channel prediction that is both adaptable to diverse scenarios and computationally feasible for real-time channel prediction.

Crucially, to avoid treating AI as a mere black-box tool and to ensure physical consistency, we integrate domain-specific propagation knowledge into the AI modeling process. Several key strategies are considered: 1) multi-modal inputs are used to construct a digital propagation scenario for AI modeling. 2) embedding physical principles are embedded into neural network architecture or loss functions to guide and constrain the learning, thus enhancing model plausibility. 3) flexible frameworks are designed to optimize accuracy-efficiency balance. 4) models are trained on multi-scenario datasets and architecture is improved to ensure robust performance with generalization and interpretability. 

The subsequent sections of this paper will elaborate on technical implementation and validation of these focused strategies. Specifically, Section II elaborates on how to extract and use physical environmental features. Section III presents design of model architecture and integration of prior propagation knowledge. Section IV investigates generalization and interpretability. Finally, Section V provides a summary of this paper.

\section{Physical Environment Representation and Extraction}
\label{sec:phys_env_rep}
Radio propagation is inherently a process of physical interaction between radio signals and the surrounding environment. 
Consequently, accuracy of AI-empowered channel modeling is highly dependent on how effectively the physical environment is encoded by neural networks. 
Selection of model input and feature representations thus play a pivotal role in modeling framework\cite{add0108-3,add0108-5}.
This section analyzes physical environment representations and investigates how the variations in multi-scale structure, geometric granularity, and environment semantic information influence channel characteristics and AI-empowered channel modeling.
\subsection{Characterization of Environmental Features}
\label{subsec:feature_char}

\begin{figure}[!t]
    \centering
    \includegraphics[width=0.95\linewidth]{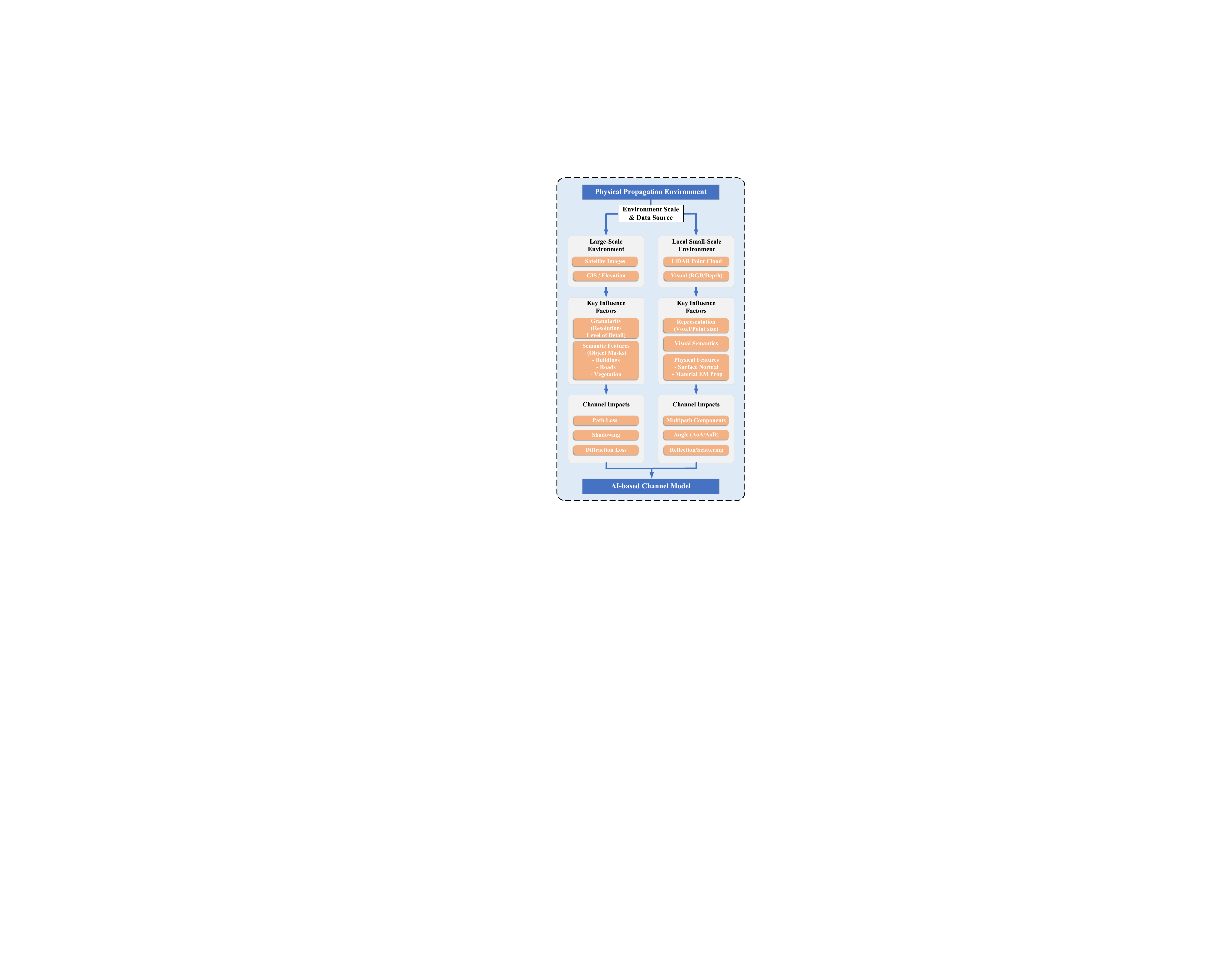}
    \caption{Hierarchical framework of environmental feature extraction for AI-based channel modeling.}
    \label{fig:env_framework}
\end{figure}

To construct a high-fidelity mapping between physical environment and wireless channel, it is essential to categorize environmental features according to the spatial extent and impact on signal propagation. The hierarchical relationship between data sources, features, and their relevance to channel modeling is illustrated in Fig. \ref{fig:env_framework}. In communication scenarios, environmental data are generally acquired through remote sensing technologies such as satellite photography, LiDAR scanning, and visual cameras, which provide information covering both large-scale environments and small-scale local scenarios.

In large-scale environments, primary data source is typically two-dimensional satellite image or geographic information system (GIS) data. The input is characterized by two critical dimensions: granularity and semantic features. 
Granularity determines spatial sampling density of environment and level of geometric detail available for feature extraction. It corresponds to the smallest spatial unit that the model can resolve.
Similarly, environment semantic features provides structural context for radio propagation by identifying key environmental elements\cite{zhang2025channel}, such as obstacles, which are directly related to diffraction and shadowing effects.

For small-scale local environments, high-precision three-dimensional representation is generally required to capture fine-grained interaction effects. Here, the input information typically consists of point clouds and panoramic visual images.
Point cloud representations encode geometric fidelity and physical surface attributes. Geometric fidelity determines accuracy of spatial reconstruction, and electromagnetic material properties of surfaces govern energy loss during reflection and transmission processes.
Panoramic visual data, including depth maps and environment semantic segmentation, provide additional spatial references.
These physical information enable AI models to directly link environmental geometry to channel impulse response.

\subsection{Environmental Granularity}
\label{subsec:impact_analysis}

Impact of environmental granularity on channel prediction depends on input representation. In large-scale environments described by two-dimensional satellite image, image resolution controls amount of usable spatial structure. Quantitative analysis in \cite{qiu2025} reveals a non-monotonic, U-shaped relationship between resolution and path loss prediction accuracy.
For the environment with low resolutions, prediction errors are dominated by critical structures, such as street layouts and building boundaries. 
As resolution increases, performance improves until reaching an optimal interval where geometric boundaries of buildings and roads are sufficiently resolved. 
However, further increasing resolution causes performance degradation. This phenomenon occurs because high resolution image introduces abundant texture noise, such as roof tiles and vegetation details. 
Thus, for channel prediction, an optimal environmental granularity exists that balances structural fidelity and texture-induced noise.

\begin{figure}[t]
    \centering
    \includegraphics[width=1.0\linewidth]{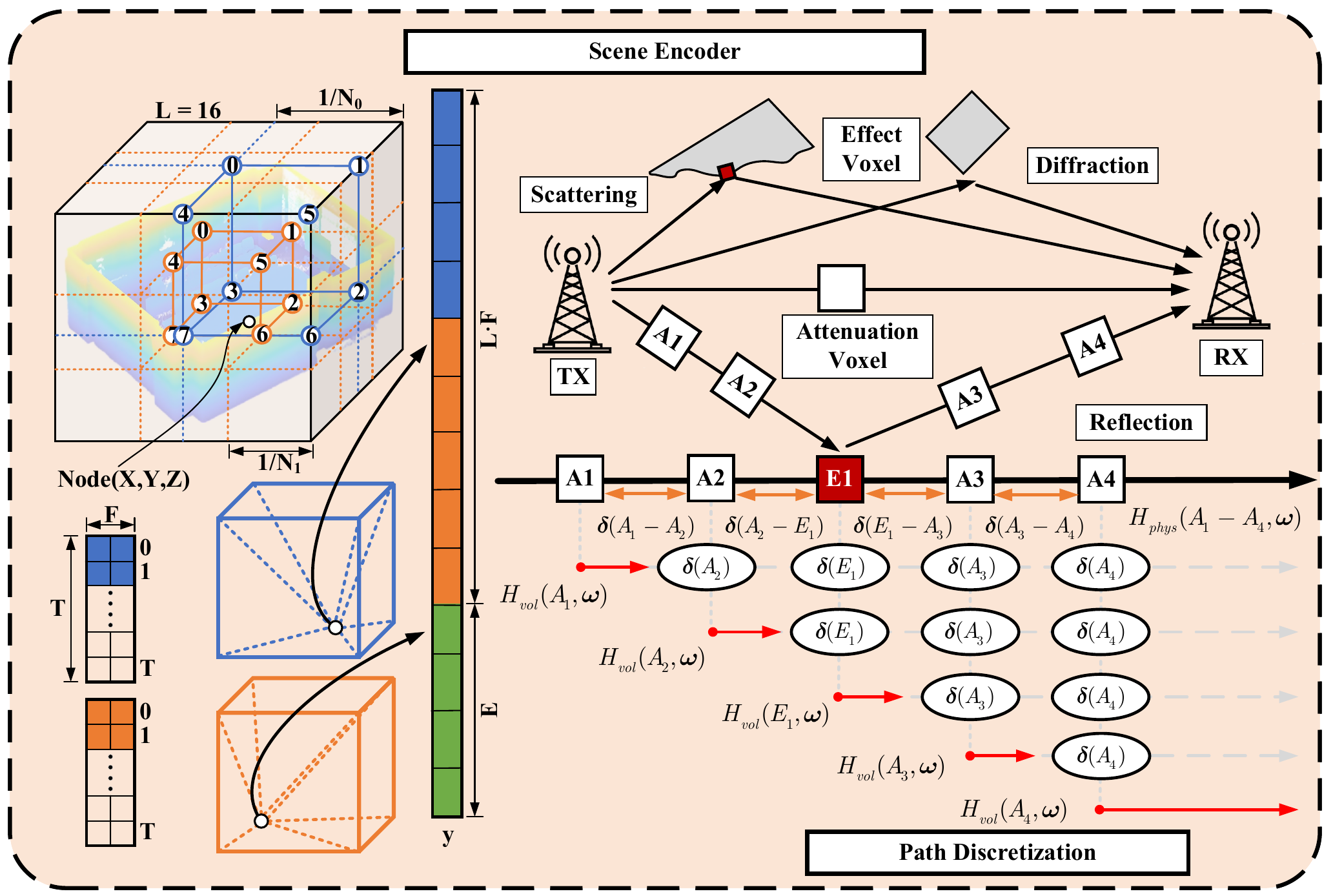} 
    \caption{A physics-inspired framework for investigating 3D environmental granularity.}
    \label{fig:fhrf_arch}
\end{figure}

\begin{figure*}[t]
    \centering
    \includegraphics[width=0.9\textwidth]{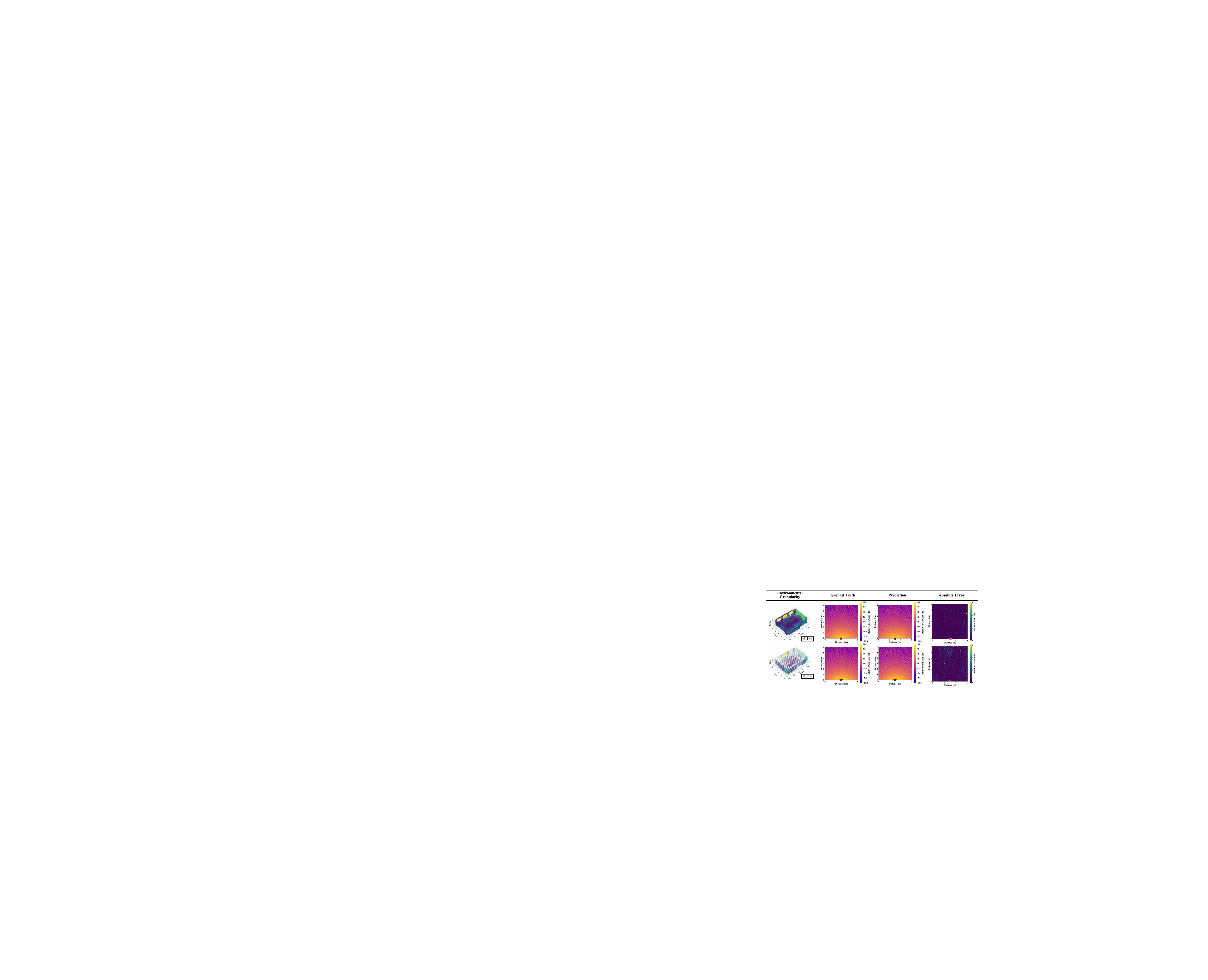} 
    \caption{Channel reconstruction error under different 3D environmental granularities. The absolute error represents the deviation between the ground truth and predicted values. Increasing voxel precision leads to a monotonic reduction in error by providing more accurate geometric priors for ray interaction modeling.}
    \label{fig:3d_combined}
\end{figure*}

In contrast, small-scale local channel modeling exhibits a fundamentally different behavior. To investigate impact of granularity, we employ a physics-inspired neural framework\cite{qi2025} as illustrated in Fig. \ref{fig:fhrf_arch}. This architecture uses voxel grids to represent small-scale environments and infers channel responses by tracing propagation paths within voxel space. Here, granularity corresponds to voxel size and point density. Experimental results in Fig. \ref{fig:3d_combined} demonstrate a strictly monotonic improvement in prediction accuracy as environmental fidelity increases. Reducing voxel size allows the path-finding algorithm to resolve detailed object contours rather than rough block approximations. High geometric fidelity ensures that surface normal vectors, which describe perpendicular orientation of geometric faces, are correctly aligned for reflection calculations. Consequently, unlike 2D case, 3D granularity directly translates to superior physical geometric priors and enhances capability of model to capture complex multipath interference patterns.

\subsection{Feature Fusion and Physics Enhancement}
\label{subsec:Feature Fusion and Physics Enhancement}
Constructing a high-fidelity mapping from physical environment to wireless channel requires effective feature fusion and physical enhancement. This process involves integrating multi-dimensional information ranging from large-scale geographic land to local visual object geometric and electromagnetic properties.

In large-scale environments, relying on raw RGB texture is often insufficient due to visual artifacts such as shadows. Incorporating explicit environmental semantic masks significantly enhances model performance. It is found in \cite{s24} that building footprint information yields the most substantial error reduction, confirming that building boundaries dominate non-line-of-sight blockage. Road information provides secondary gains for signal energy, whereas vegetation features contribute marginally. Effective feature fusion relies on selecting structural elements that define propagation geometry rather than aggregating irrelevant semantic categories.

\begin{figure}[t]
    \centering
    \includegraphics[width=\linewidth]{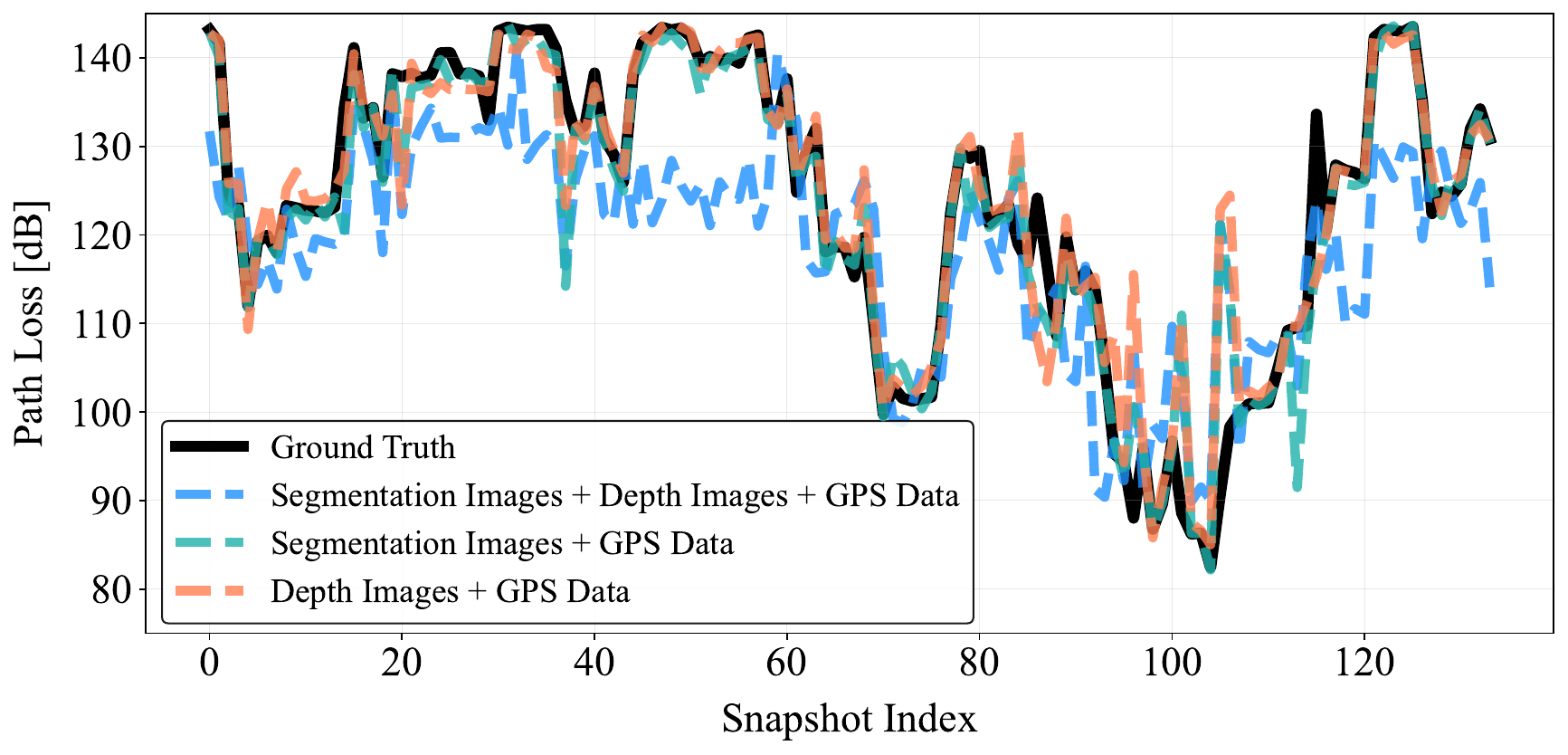} 
    \caption{Path loss prediction performance using visual sensing.}
    \label{fig:visual_pl}
\end{figure}

In small-scale environments, achieving precise channel prediction requires integrating fine-grained small-scale features, including object semantics, depth information, and specific physical environmental attributes. Consequently, fusion of heterogeneous inputs in local scenarios requires careful alignment. We investigate the impact of augmenting visual data with environmental semantic segmentation and depth estimation, using the measurement data in \cite{11142292}. As shown in Fig.\ref{fig:visual_pl}, prediction root mean square errors (RMSE) using combined segmentation and depth images, segmentation images only, and depth images only are 10.67 dB, 5.31 dB, and 6.05 dB, respectively. The segmentation-only approach achieves the lowest error, outperforming the combined input method by 5.36 dB. This performance degradation in combined approach highlights a critical challenge known as feature misalignment, where segmentation and depth maps derived from different algorithms suffer from spatial inconsistencies or modal noise. Without specialized alignment mechanisms, conflicting features confuse the neural network. This suggests that for visual sensing, feature coherence is more critical than quantity of modalities.

\begin{figure}[t]
    \centering
    \includegraphics[width=\linewidth]{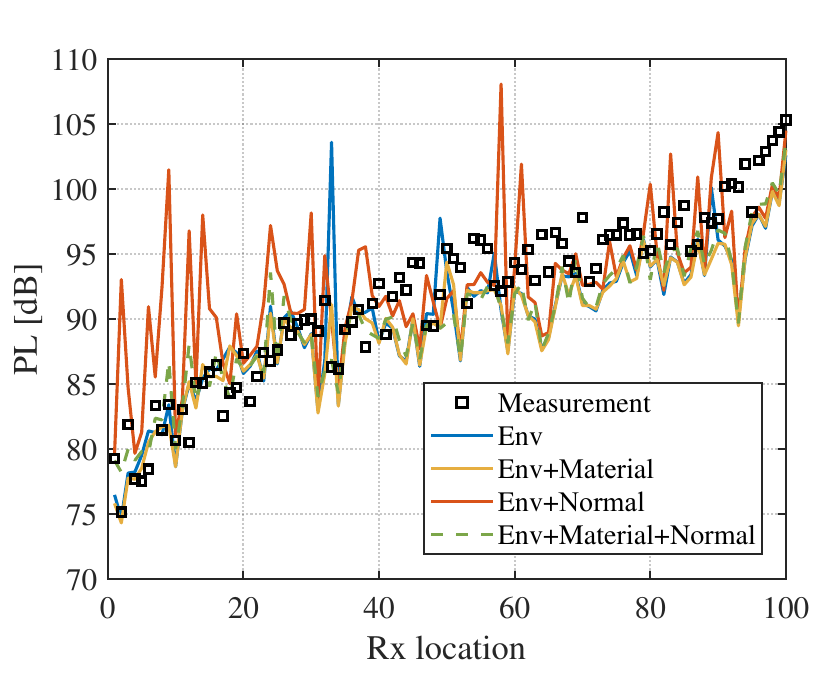} 
    \caption{Impact of fusing 3D physical information on channel prediction. Material: Material properties, including conductivity, relative permittivity, and scattering coefficient of the surface. Normal: Normal vector used to describe the orientation of scene surface. The normal vector, together with incident angle, determines the outgoing directions of reflected and scattered rays.}
    \label{fig:3d_material_pl}
\end{figure}

Furthermore, channel prediction accuracy can be improved by physics enhancement. For large-scale environments, it is feasible to integrate Fresnel zone clearance to better identify line-of-sight obstructions, and heuristic diffraction models based on building height profiles can be used to estimate shadow fading. Regarding local small-scale environments, beyond basic object semantics and depth, physical representation can be enriched by introducing theoretical Radar Cross Section priors to quantify scattering intensity of distinct objects. Moreover, leveraging precise representation of scenario information allows for explicit incorporation of intrinsic physical material properties for both environmental background and distinct semantic objects, including conductivity, relative permittivity, and scattering coefficients, alongside surface normals, thereby embedding physical propagation mechanisms into feature space. In terms of electromagnetic waves propagation mechanisms, surface normal vectors together with incident angle, determines outgoing directions of reflected and scattered rays. Material information characterizes intrinsic electromagnetic properties of interaction surfaces and affects multipath attenuation. Motivated by these physical insights, different combinations of environmental features can be designed as network inputs to evaluate the contributions to prediction performance. Specifically, we evaluate the consistency between predictions and measured results under different physics enhancement using a point cloud-based network. The network is composed of two stages. First, PointNet++ is employed to encode point cloud and extract environmental features. Subsequently, a Transformer-based model is used to predict ray propagation directions and multipath attenuation, enabling modeling of path propagation behavior. A real-world 3D point cloud and the corresponding channel measurement data used for network validation are acquired through an on-site measurement campaign conducted at a typical indoor entrance hall. Channel measurements are carried out at 28 GHz with a bandwidth of 1 GHz. Channel frequency responses over 1024 OFDM subcarriers are captured at fixed positions from a single-antenna omnidirectional transmitter to a planar $4 ‌\times 8$ antenna array. As shown in Fig. \ref{fig:3d_material_pl}, geometry-only input captures overall path loss trend and misses local fluctuations. Incorporating material electromagnetic properties reduces RMSE by approximately 1.89 dB. In contrast, explicitly including surface normals provides limited marginal gains or even introduces redundancy. This implies that the neural networks can implicitly learn orientation features from high-fidelity 3D structure, whereas material attributes act as independent electromagnetic boundary conditions playing a decisive role in energy modeling.

\subsection{Mapping Environmental Features to Propagation Mechanisms}
\label{subsec:synthesis}
The above results show that effects of using environmental features depends not only on data resolution. It is more determined by the consistency with underlying electromagnetic propagation mechanisms. Accordingly, environmental feature extraction and fusion should move from generic data processing toward a physics-aware mapping strategy.

In large-scale scenarios, interaction between radio waves and environment is dominated by shadowing and large-scale diffraction. Our analysis demonstrates that raw visual textures in optical imagery often act as high frequency noise that confuses neural networks. Therefore, the optimal representation is not a high resolution image, and it should be a deterministic semantic structure. By filtering out textural details and retaining only footprints of buildings and road networks, we explicitly assist model in identifying diffraction edges and obstacles responsible for signal blockage.

Conversely, small-scale modeling demands a different approach governed by ray optical behaviors. In this regime, environmental granularity equates to fidelity of physical simulation. Geometry defines the existence of propagation paths, while material properties govern energy gain through reflection and scattering. This implies that the environment functions as a complex electromagnetic boundary condition rather than a simple geometric obstruction. Thus, incorporating multidimensional physical attributes is indispensable for accurate channel prediction. Ultimately, effective AI-driven channel modeling depends on representing environmental data with physically meaningful features, allowing neural networks to learn propagation physics rather than spurious correlations.

\section{AI-Empowered Channel Inference}

In this paper, we introduce the concept of channel inference, which can be considered as AI-empowered site-specific channel prediction. Channel inference aims to rapidly predict and continuously update wireless channel states under known or perceptible environmental conditions, enabling site-specific link-level or area-level (i.e., radio map) channel prediction. Unlike conventional channel simulation approaches that mainly rely on statistical assumptions or offline geometric modeling, channel inference emphasizes accurate alignment with real propagation environments and real-time awareness of spatial–temporal channel evolution. In this paradigm, wireless channel is treated as a functional mapping of environment state, driven jointly by scenario geometry, object distribution, and dynamic behaviors. Therefore, channel inference requires high modeling accuracy and low computational latency to support real-time applications. Given the strong nonlinearity, non-stationarity, and environment dependency of radio propagation, purely analytical or statistical models struggle to achieve both accuracy and efficiency, making AI
based methods an essential choice.

\subsection{Channel Inference Paradigm}

\begin{figure*}[t]
    \centering
    \includegraphics[width=6.6in]{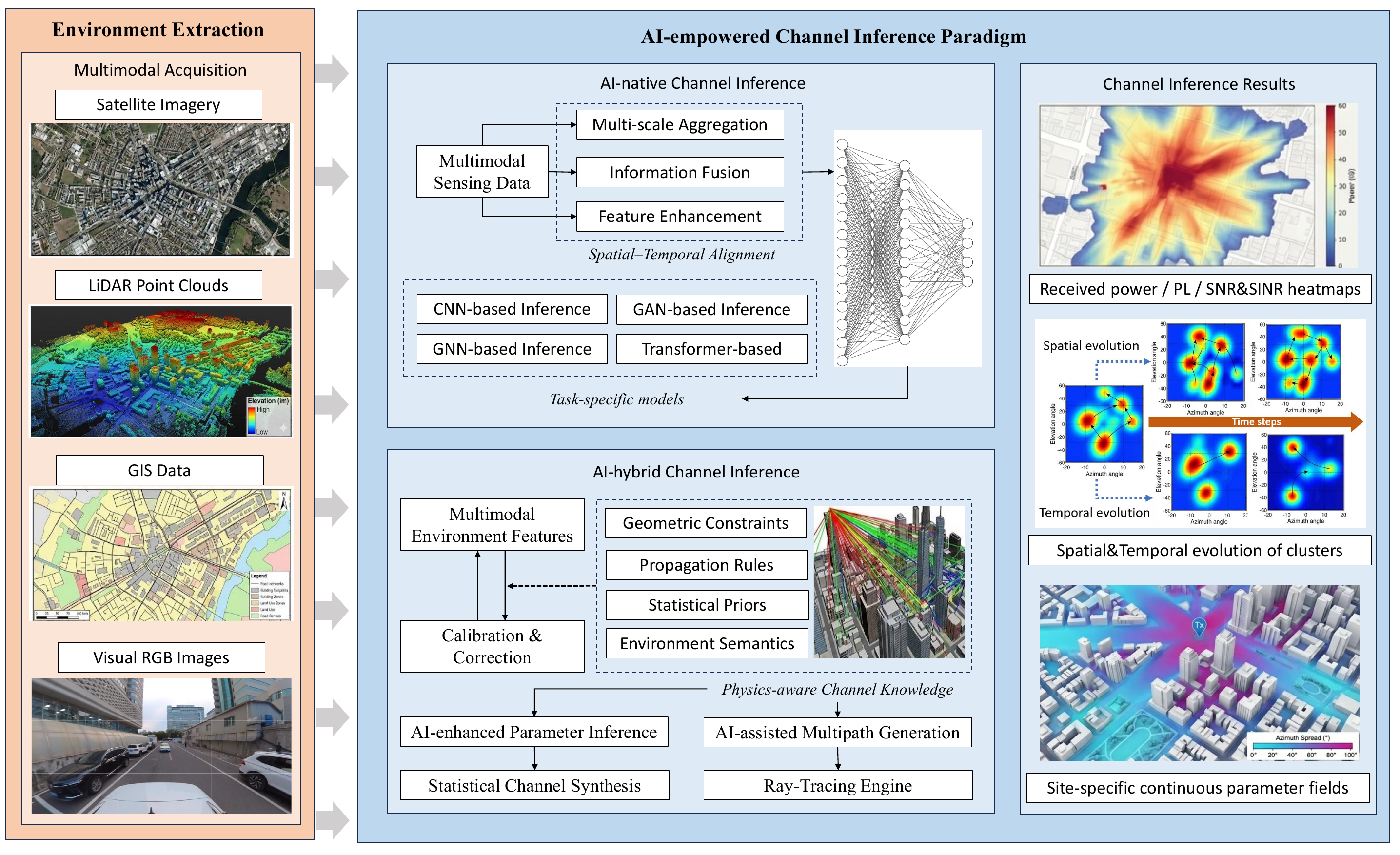} 
    \caption{Paradigm of AI-empowered channel inference.}
    \label{fig:overview}
\end{figure*}

We illustrate the paradigm of AI-empowered channel inference in Fig.~\ref{fig:overview}. The foundation of channel inference lies in effective environmental feature representation. Through multimodal environment acquisition, wireless propagation scenarios can be characterized from multiple perspectives and scales, including large scale geographic layouts, three-dimensional structures, local blockages, and environmental semantic information. Typical inputs include satellite image, GIS data, LiDAR point clouds, and visual images, which differ significantly in resolution, scale, and physical meaning. These heterogeneous data sources require unified feature extraction and fusion mechanism. The goal of environment representation is not to fully reconstruct physical scenario. It aims to generate compact and propagation-relevant feature representations that
can be efficiently exploited by channel inference models.

Based on environment representation, AI-empowered channel inference establishes a unified and flexible inference framework that supports both AI-native and AI-hybrid inference approaches. For the AI-native paradigm, neural networks directly learn the mapping from environmental features to channel states. Traditional propagation mechanisms or path-level calculations are not explicitly modeled. Instead, implicit relationships between environment and channel responses are learned from data. Depending on inference objective, task-specific models can be designed to predict path loss, angle and delay distributions, multipath cluster structures, or instantaneous channel responses. Recent advances in convolutional neural networks\cite{wang2024channel}, generative adversarial network\cite{zhang2023generative}, graph neural networks\cite{10924577}, and Transformer-based models\cite{10965849} have shown strong capabilities in modeling spatial structures and temporal evolution. Moreover, with the emergence of large language models, AI-native inference is evolving toward unified inference models that support multiple channel prediction tasks and cross-scenario generalization. This paradigm offers end-to-end inference, high computational efficiency, and ease of deployment, making it suitable for latency-sensitive applications.

In contrast, the AI-hybrid channel inference paradigm focuses on collaboration between data-driven models and traditional physical or statistical channel models. Rather than fully replacing physics-based modeling, AI is integrated as an auxiliary or enhancement module by incorporating propagation rules, geometric constraints, statistical priors, and environment semantics into inference process. On one hand, neural networks can assist deterministic models by enabling path selection, multipath parameter correction, or propagation decision-making in complex scenarios, thereby reducing computational complexity and improving robustness. On the other hand, AI can enhance statistical channel modeling by learning mappings between environment features and statistical parameters, enabling scenario-adaptive parameter generation. This hybrid approach preserves physical consistency and interpretability while leveraging expressive power of data-driven models, achieving a balanced trade-off among accuracy, generalization, and efficiency.

The outputs of channel inference range from large-scale parameters to fine-grained multipath characteristics. At the parameter level, inference results may include spatial distributions of received power, path loss, SNR, or SINR, which describe coverage and link quality. At a finer level, temporal evolution of multipath clusters and spatial distribution of continuous channel parameter fields can also be obtained. These inferred channel representations directly support communication system design and provide valuable channel information for network optimization in complex environments.

\subsection{Model Architecture}

\begin{figure}
    \centering
    \includegraphics[width=.5\textwidth]{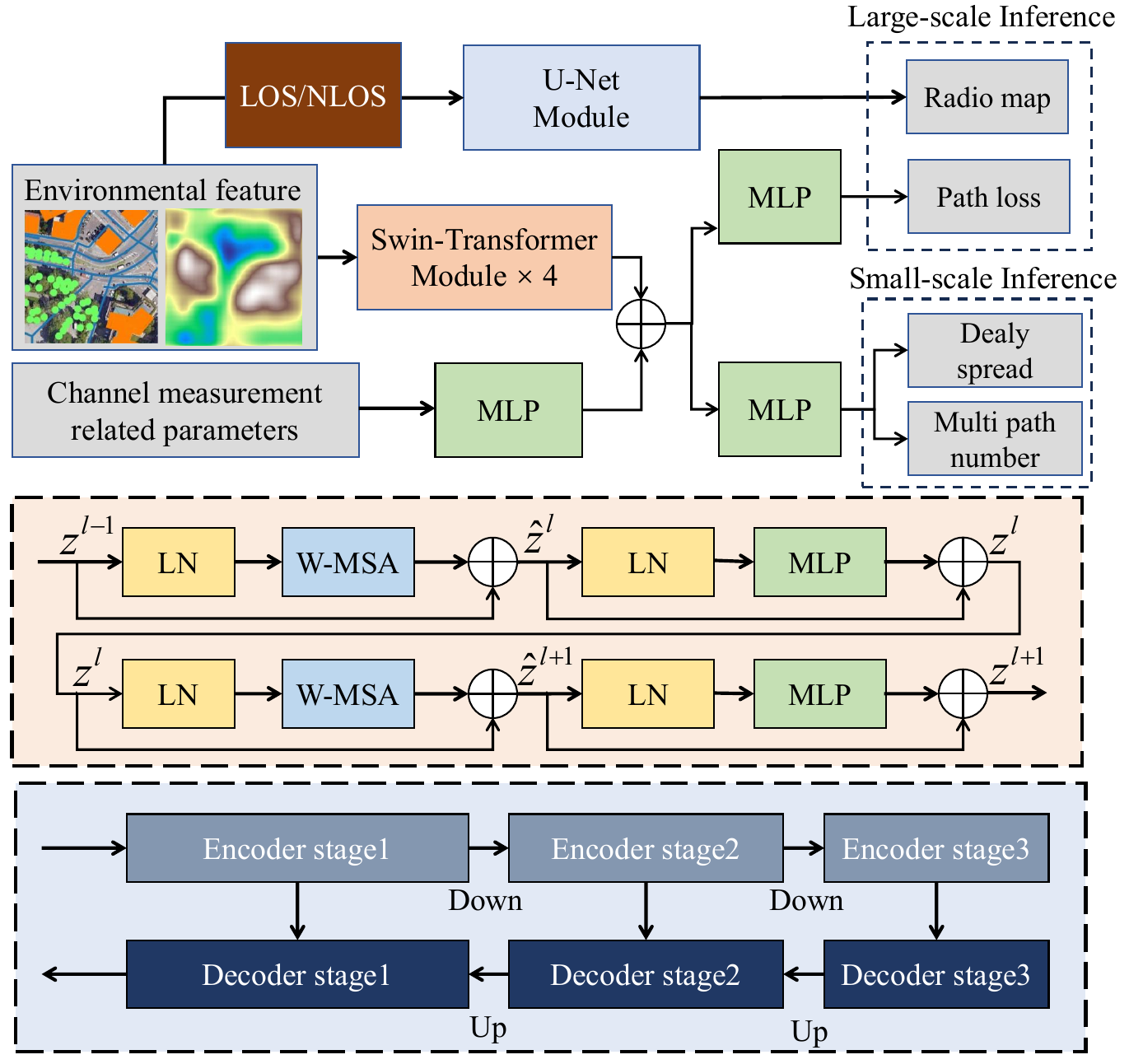}
    \caption{Model architecture of AI-empowered channel inference. LN denotes Layer Normalization, which is used to stabilize training and accelerate convergence. W-MSA denotes Window-based Multi-Head Self-Attention, where self-attention is computed within local non-overlapping windows to reduce computational complexity.}
    \label{framework}
\end{figure}

In this subsection, we present an AI-based channel inference architecture, as illustrated in Fig. \ref{framework}, which aims to accurately infer both large- and small-scale channel parameters. The architecture integrates U-Net, Swin-Transformer, and Multi Layer Perceptron (MLP) modules into a unified framework that supports hierarchical feature encoding, decoding reconstruction, and dual scale parameter estimation.
The model takes two complementary data modalities as input. Environmental features provide macro level information such as terrain characteristics, obstacle distribution, and propagation environment categories, whereas channel measurements capture signal level observations including received signal strength and phase variations. These heterogeneous inputs are jointly exploited to enhance inference robustness and physical consistency.

For the core of framework, a Swin-Transformer encoder composed of three successive encoding stages, namely Encoder Stage 1 to Encoder Stage 3, performs hierarchical feature extraction \cite{liu2021swin}. 
Each stage adopts a standardized structure consisting of Layer Normalization (LN), Window Multi Head Self Attention (W-MSA), a second LN, and a MLP, with residual connections to facilitate stable training and long range dependency modeling. Downsampling operations between adjacent stages progressively compress feature dimensions and promote higher level abstraction. Correspondingly, three decoding stages reconstruct the encoded representations back to the parameter related feature space, enabling subsequent dual scale inference.
To enhance spatial feature modeling, it incorporates a U-Net based radio map module that exploits encoder decoder symmetry and skip connections to capture spatial propagation characteristics. In addition, several repeatedly deployed general purpose modules together with multiple independent MLP layers are employed for feature refinement and parameter mapping.

Large-scale channel parameters are inferred through targeted feature extraction that emphasizes large-scale propagation behavior. The outputs include line of sight and non line of sight (LOS/NLOS) status and path loss. Their inference primarily relies on the Swin-Transformer encoder for environmental feature mining and the U-Net radio map module for spatial propagation modeling.
In contrast, small-scale channel parameters are estimated through refined feature decoding and non linear mapping via MLP layers. These parameters include delay spread, multipath related amplitude and phase correlation, and the number of effective multipath components. Their inference depends on the decoder’s ability to restore fine grained features and strong non-linear representation capability of MLPs, enabling accurate modeling of small-scale channel fluctuations.

\subsection{Channel Inference Results}
\begin{figure*}
    \centering
    \includegraphics[width=.95\textwidth]{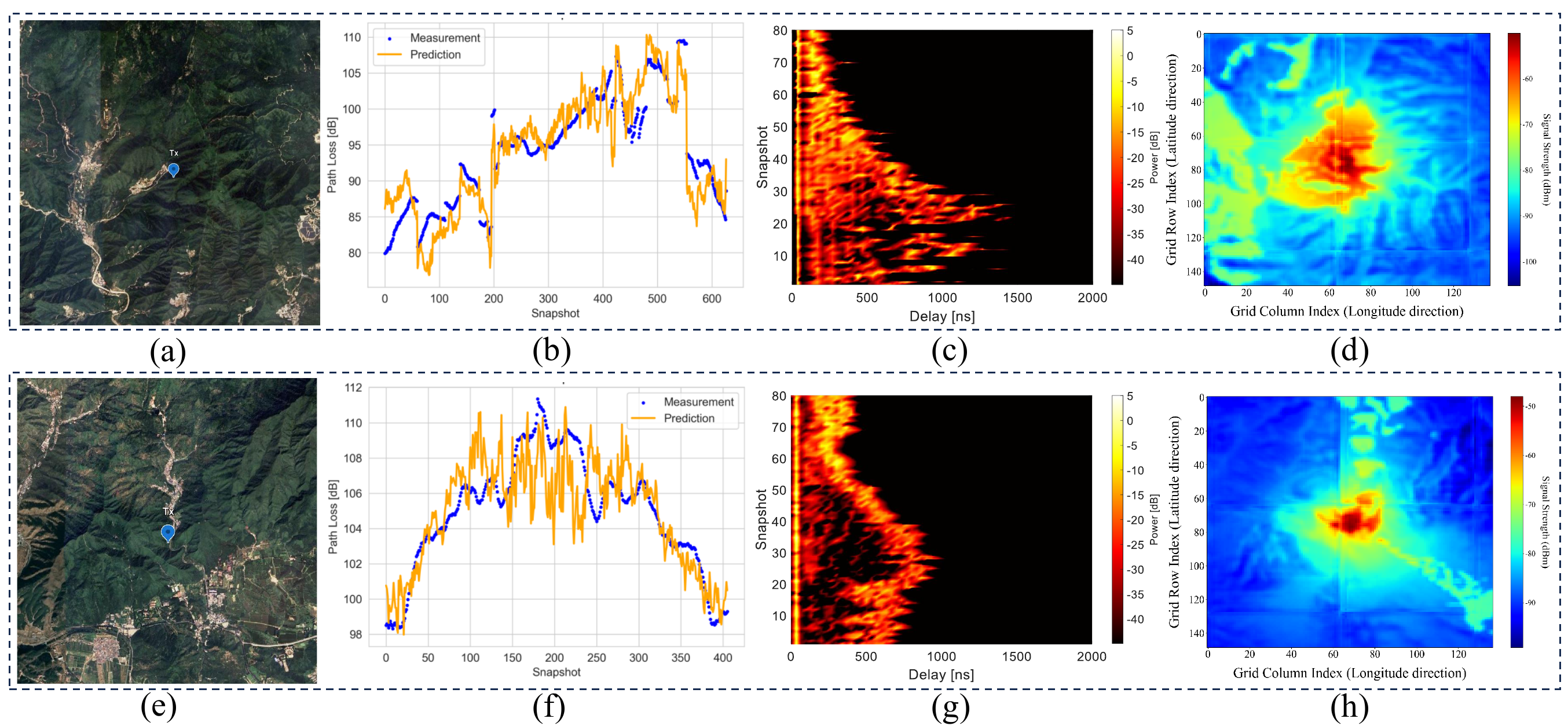}
    \caption{Performance evaluation of the proposed AI-based channel inference framework.
    (a)-(d) are for Scenario 1, and (e)-(g) are for Scenario 2.
    (a) and (e): environmental features, 
    (b) and (f): path loss prediction results, 
    (c) and (g): PDP prediction results, 
    and (d) and (h): radio map prediction results.}
    \label{results_v2}
\end{figure*}

To validate the proposed architecture, extensive channel measurement campaigns are conducted in suburban and urban environments at a carrier frequency of 5.9 GHz with a bandwidth of 30 MHz \cite{10840926}. 
The transmitter locations and receiver trajectory configurations vary across different measurement scenarios, resulting in a large-scale dataset suitable for both model training and evaluation. 
Nearly 100,000 data sets are obtained across 20 different scenarios.
Two representative scenarios are selected as the test set, with the corresponding transmitter locations illustrated in Figs. \ref{results_v2}(a) and (e). 
Measurement data from the remaining scenarios are used for training, with the inference results for Scenario 1 and Scenario 2 shown in Figs. \ref{results_v2}(a)–(d) and (e)–(h), respectively.

Satellite images corresponding to the two geographical environments are shown in Figs. \ref{results_v2}(a) and (e), which serves as environmental input to the proposed framework. The two scenarios exhibit distinct terrain characteristics and obstacle distributions, thereby providing diverse large-scale propagation conditions for subsequent evaluation.
Path loss prediction results along the measurement trajectories are presented in Figs. \ref{results_v2}(b) and (f). For both scenarios, the predicted path loss demonstrates strong agreement with the measured data, accurately capturing the overall attenuation trend as well as location-dependent variations. This consistency indicates that the Swin-Transformer encoder effectively extracts high-level environmental semantics and models large-scale propagation behavior across heterogeneous terrain conditions.

Figs. \ref{results_v2}(c) and (g) depict the predicted power delay profile (PDP). The results demonstrate that the proposed model successfully reproduces the dominant multipath components and the temporal evolution. In both datasets, the predicted PDPs exhibit pronounced energy concentration at early delays, followed by gradually decaying multipath components, which is consistent with underlying physical propagation mechanisms. These observations confirm that the decoder and MLP-based nonlinear mapping can effectively recover fine-grained features and capture the stochastic characteristics of small-scale channel fluctuations.

The wide-area radio map prediction results are shown in Figs. \ref{results_v2}(d) and (h), which further demonstrate generalization capability of the proposed framework. The predicted received signal strength distributions exhibit smooth spatial transitions and well-defined coverage patterns that align closely with the surrounding terrain and obstacle layouts. In particular, regions with strong and weak signal strength correspond to plausible line-of-sight and obstructed propagation conditions, respectively, indicating that the U-Net-based radio map module effectively integrates spatial contextual information with learned propagation features.

Overall, the results from Scenario 1 and 2 consistently demonstrate that the proposed architecture enables accurate inference of large- and small-scale channel parameters, and simultaneously supports wide-area radio map prediction. The strong cross-scenario consistency validates the robustness and generalization capability of the proposed channel inference framework.

\section{Generalization and Interpretability}

Generalization and interpretability are the fundamental basis that determines whether models can be reliably deployed in practical wireless systems. Generalization refers to model ability to maintain accurate predictions in different scenarios and propagation conditions. Its core objective is to prevent overfitting to localized environmental features, thereby enabling representation of the underlying propagation mechanisms. Interpretability focuses on consistency between model internal decision-making process and the physical laws of radio propagation. It aims to explain how environmental features are mapped to variations in channel parameters through specific propagation mechanisms. This section presents approaches for improving cross-scenario generalization and interpretable design strategies that incorporate propagation knowledge into AI-based model architectures.

\subsection{Model Generalization}

Wireless propagation environments exhibit high variability in terrain and object structures. Training data cannot cover all possible spatial distributions. We present some potential strategies to enhance model generalization, with a focus on co-design of training and learning processes.

The first step toward improving generalization is construction of large-scale, multimodal datasets that capture diversity of real-world propagation environments. Traditional datasets relying solely on geographic coordinates and path-loss labels typically fail to provide sufficient propagation diversities. By incorporating environment-aware information such as satellite image, building footprint vectors, and terrain elevation data, model can learn visual and structural constraints. Multi-scenario training datasets should be constructed, and this allows the model to have wider variety of object combinations, blockage relationship, and statistical propagation characteristics during training process, thereby reducing its dependence on geometric features of any specific region.

In pre-training stage, model is constrained using large-scale heterogeneous environmental data. This guides hidden-layer features to gradually extract stable structures that are relevant to propagation. In transfer learning stage, a small number of samples from the target domain can be used to recalibrate high-level mappings, enabling rapid alignment across different environments within a relatively fixed feature space \cite{wang2021transfer}. In practice, pre-training typically uses simulated data to learn intrinsic correlations among building contours, street orientations, and blockage structures. The low-level feature extraction network is frozen, and only the high-level channel-mapping layers are updated. This allows the model to retain general environment representations while rapidly adapting to new propagation statistics.

When the goal is extended from path loss prediction to multi-task learning, the model needs to learn the inherent structural relationships in propagation process. This requires higher robustness to environmental variations. Beyond data and model design, formulation of tasks directly affects model generalization capability. In addition, decomposing complex channel mapping tasks into multiple physically meaningful subtasks, such as environmental structure analysis and propagation mechanism inference, encourages the model to capture shared propagation structure features, forming strongly coupled expressions of physical laws across different tasks.

Following the above idea, we train the model based on Swin Transformer to improve generalization with pre-training and transfer learning, as illustrated in Fig. \ref{framework}. The model adopts a dual-branch feature fusion architecture together with a multi-head output design, in which channel parameters are integrated with high-resolution satellite image and digital elevation maps around the receiver.
The model is trained using measured and simulated data collected from multiple scenarios, covering a wide variety of geographical environments and channel conditions. However, due to the inherent complexity and high cost of channel measurements, the amount of real-world data remains limited and cannot fully span the diversity of practical environments. To address this limitation, large-scale simulation data generated from digital elevation maps are used for pre-training, allowing the model to learn diverse environment-aware feature representations related to channel characteristics. Subsequently, real-world measurement data are employed for fine-tuning, further improving model performance.
Specifically, during the fine-tuning stage, the weights of the low-level feature extractor in the pre-trained model are kept fixed, and only the parameters of high-level prediction network are updated. This optimization process can be formulated as:
\begin{equation}
\min_{\theta_h} \; \mathcal{L}\Big(z\big(f(x;\theta_f),\,\theta_h\big),\, y\Big)
\label{eq:finetune}
\end{equation}
where $\mathcal{L}(\cdot)$ denotes loss function used in fine-tuning stage, 
$z(\cdot)$ represents model output, 
$y$ denotes ground-truth label, 
$f(\cdot)$ is low-level feature extractor, 
$x$ represents input environmental features, 
$\theta_f$ denotes parameters of feature extractor, 
and $\theta_h$ denotes parameters of high-level prediction network.
It is found that the proposed model achieves an RMSE of 4.71 dB in path loss prediction under suburban scenarios covering 30 MHz$\textendash$5.9 GHz. Compared with training from scratch, adoption of a transfer learning strategy reduces training time in target scenario by 60\%$\textendash$75\%. Moreover, model generalization capability in long-distance link is significantly improved, yielding an average 3.62 dB reduction in RMSE of path-loss prediction, which demonstrates the proposed multi-dimensional generalization enhancement strategy.

\subsection{Model Interpretability}

In AI-based channel modeling, priori propagation knowledge can help model better understand the underlying physical characteristics in data. Such propagation knowledge can well improve interpretability for AI models. Incorporating propagation knowledge into dataset construction by fully augmenting input data, such as environmental images and building density, is critical for improving model performance. Proper feature design aims to represent known, possibly non-linear, interactions between input and output that the model needs to learn from data, which can enhance both prediction accuracy and interpretability. 

\begin{figure}
\centerline{\includegraphics[width=3.5in]{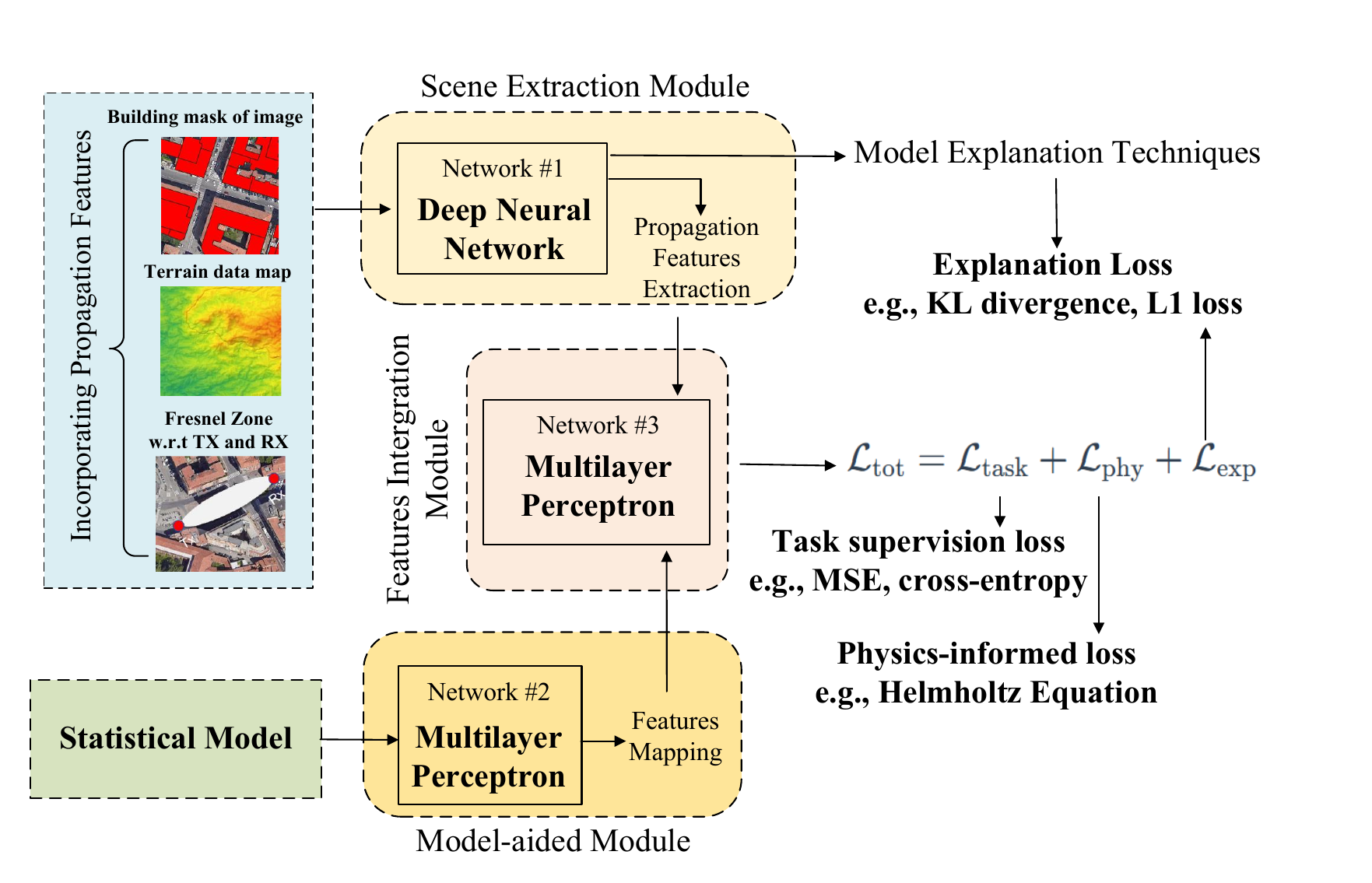}}
\caption{Illustration of AI techniques incorporating propagation knowledge for channel modeling: dataset construction, model architectures, and loss function design.\label{fig1}}
\end{figure}

Propagation knowledge can be directly incorporated into AI-based architecture through model-aided design, achieving a model-driven rather than purely data-driven approach. AI model predictions, which use automated feature extraction from satellite images aided by a simple statistical model, can be aligned with relevant physical knowledge. As an illustrative example, in Fig. \ref{fig1}, for a path loss regression problem, it can be useful to choose a simple path loss model to augment DNN model. Another effective strategy is to incorporate physics-informed regularization terms into loss function. The AI model could be trained by minimizing a standard loss function combined with a physics-informed loss containing physical mechanism. The additional loss incorporates priori known physical laws and is compatible with model training algorithm. For example, incorporating differential equations from electromagnetic theory or laws from geometrical optics as an additional loss term  could enhance model interpretability. Explainable AI techniques can also provide saliency maps to understand which environmental features in a sample are the most responsible for the model prediction. The additional supervision signals can be sufficient to improve the model performance. In Fig. \ref{fig1}, we illustrate that the total loss could be decomposed into a task supervision loss, a physics-informed loss, and an explanation supervision loss.

Quantitative interpretability metrics can also be used to assess whether model are valid and reliable, and they typically include measures such as faithfulness \cite{samek2016evaluating}. Statistical evaluation of explanation faithfulness is often performed using perturbation-based metrics. One representative example is comprehensiveness metric. It measures the change in the prediction of model when the feature or rationale identified as important by the model explanation is occluded or masked from the original input. Visualization-based interpretability methods help reveal which parts of the input data are important and what factors contribute to its predictions. Among these methods, feature-importance visualization techniques, such as feature ranking lists, can well illustrate influence of different input features on model’s output. Image-importance visualization methods are also commonly used for interpreting deep learning models. These methods aim to reveal why a model produces a certain prediction in image-related tasks and help users understand the basis of its decisions. 

\begin{figure}
\centerline{\includegraphics[width=3.5in]{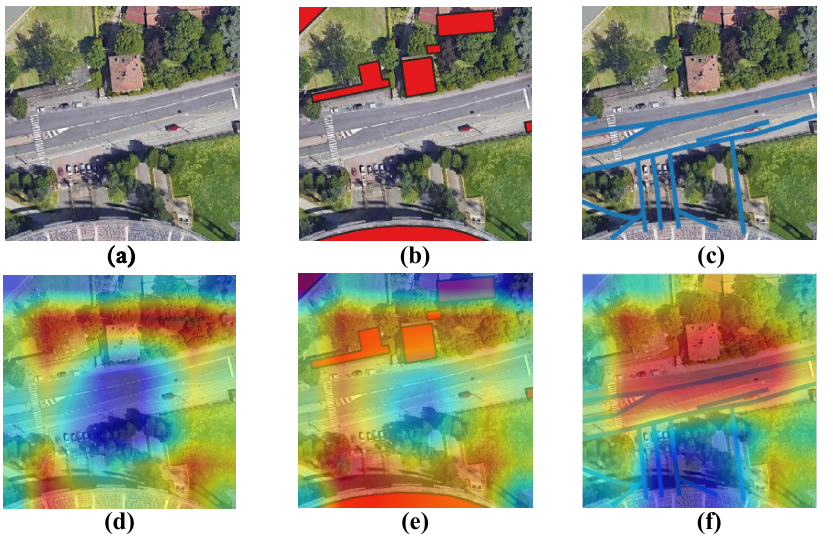}}
\caption{An illustration of images and the corresponding saliency maps for a test scenario. (a) Original satellite image; (b) building features annotated as masks; (c) road features annotated as masks; (d) saliency map generated by the baseline image; (e) saliency map generated by the model incorporating building annotations; (f) saliency map generated by the model incorporating road annotations. \label{fig2:six_subfigs}}
\end{figure}

\begin{figure}
\centerline{\includegraphics[width=3.0in]{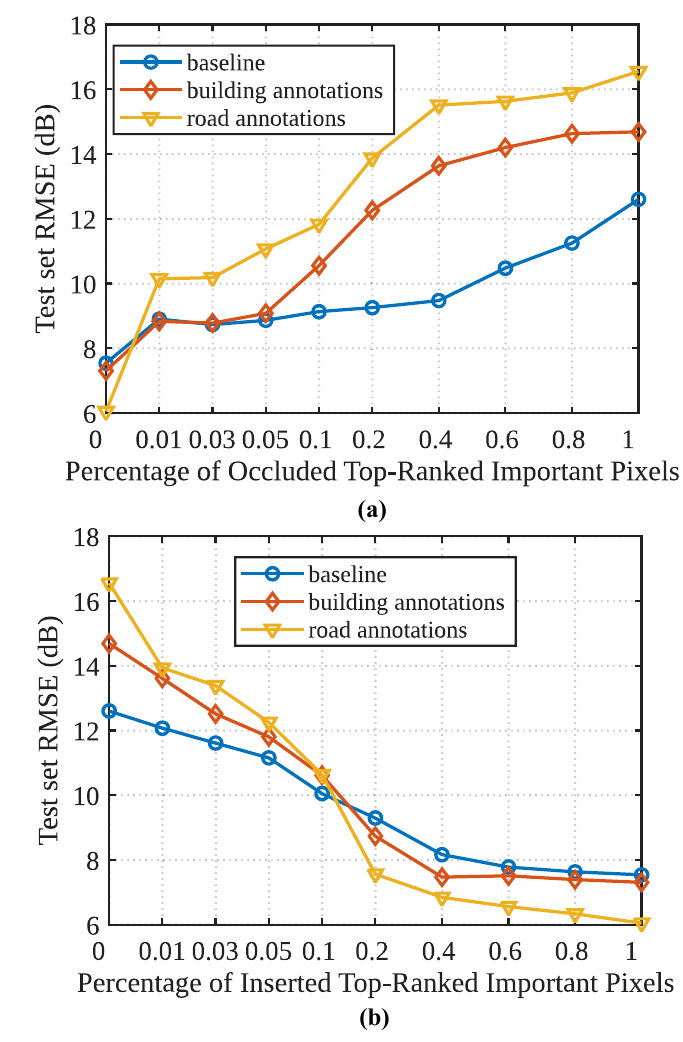}}
\caption{Perturbation analysis for the path loss prediction. (a) Occlusion perturbation results; (b) Insertion perturbation results.\label{fig1:two_subfigs}}
\end{figure}

As an illustrative example, we consider a path loss prediction task based on urban dataset in \cite{gozalvez2012ieee}. This dataset comprises coordinates of transmitters, receivers, received signal strength indicator, and other parameters collected during measurements. We select a subset of measurement data and extract the corresponding satellite images based on RX coordinates to construct dataset for our target task. To incorporate relevant propagation knowledge, road and building features are annotated separately as masks on the satellite images in each subset. We train a CNN to predict path loss given an input image along with the associated measurement parameters. The modified Grad-CAM \cite{selvaraju2017grad} is applied to generate pixel-level saliency maps for each test sample. In Fig. \ref{fig2:six_subfigs}, it is shown that incorporating diverse types of environmental features enables models to capture additional characteristics that affect radio propagation. Based on the identified important regions, pixels are ranked in descending order to identify the most influential regions. We use a perturbation method \cite{samek2016evaluating} to evaluate effect of propagation knowledge on models predictions and its faithfulness. 

In Fig. \ref{fig1:two_subfigs}, we present the results of deletion and insertion metrics \cite{petsiuk2018rise} for baseline and the models that incorporate propagation knowledge. In occlusion-based perturbation, the top-ranked important pixels are progressively occluded, and the corresponding changes in prediction performance are recorded. A faithful explanation is expected to induce rapid degradation in model performance when a small proportion of the most important pixels is occluded. As the percentage of important pixels occluded in the top ranks increases, average test RMSE of all models exhibits an overall increasing trend, indicating a degradation in prediction performance. Models incorporating propagation knowledge demonstrate a significantly steeper increase in average test RMSE compared to baseline. In particular, the model incorporating road features shows the most rapid performance degradation under the same occlusion ratio, followed by the model incorporating building features. In insertion-based perturbation, the most important pixels are gradually inserted into a fully perturbed input image. Consistent with the occlusion analysis, models incorporating propagation knowledge exhibit faster recovery in prediction performance. The model incorporating road features achieves the most rapid RMSE reduction, indicating that inserting a small proportion of highly ranked pixels is sufficient to recover most of the predictive capability. It is evident that the models incorporating propagation knowledge have better performance and interpretability than the baseline given the same training budget. 

We further extract building features as single-channel masks representing explicit environmental knowledge. Following the principle of Explanation-Guided Learning \cite{li2018tell}, these feature masks are treated as region-level importance priors, and the saliency maps generated by the model can be considered as constraints. Those saliency maps are used as feedback (i.e., in terms of loss function as illustrated in Fig. \ref{fig1}) to the AI network during training to guide the network to focus on propagation-relevant regions of environment. With this enhanced training strategy, average test RMSE of the model incorporating building-related propagation knowledge is further reduced from 7.30 dB to 7.07 dB, demonstrating the effectiveness of explanation-guided supervision, which can form a positive feedback mechanism that further improves model performance.

\section{Conclusion}
This paper establishes a foundational AI-empowered paradigm for future propagation channel modeling. By deeply extracting environment features, fusing physical propagation mechanisms, and constructing hybrid model framework, it significantly improves in terms of accuracy, generalization, and efficiency. The proposed methodologies, validated by improved quantitative metrics such as prediction RMSE and training efficiency, pave the way for developing scalable, interpretable, and real-time channel prediction systems for 6G and beyond.

\bibliographystyle{IEEEtran}

\bibliography{refs}
 
\begin{IEEEbiography}[{\includegraphics[width=1in,height=1.25in,clip,keepaspectratio]{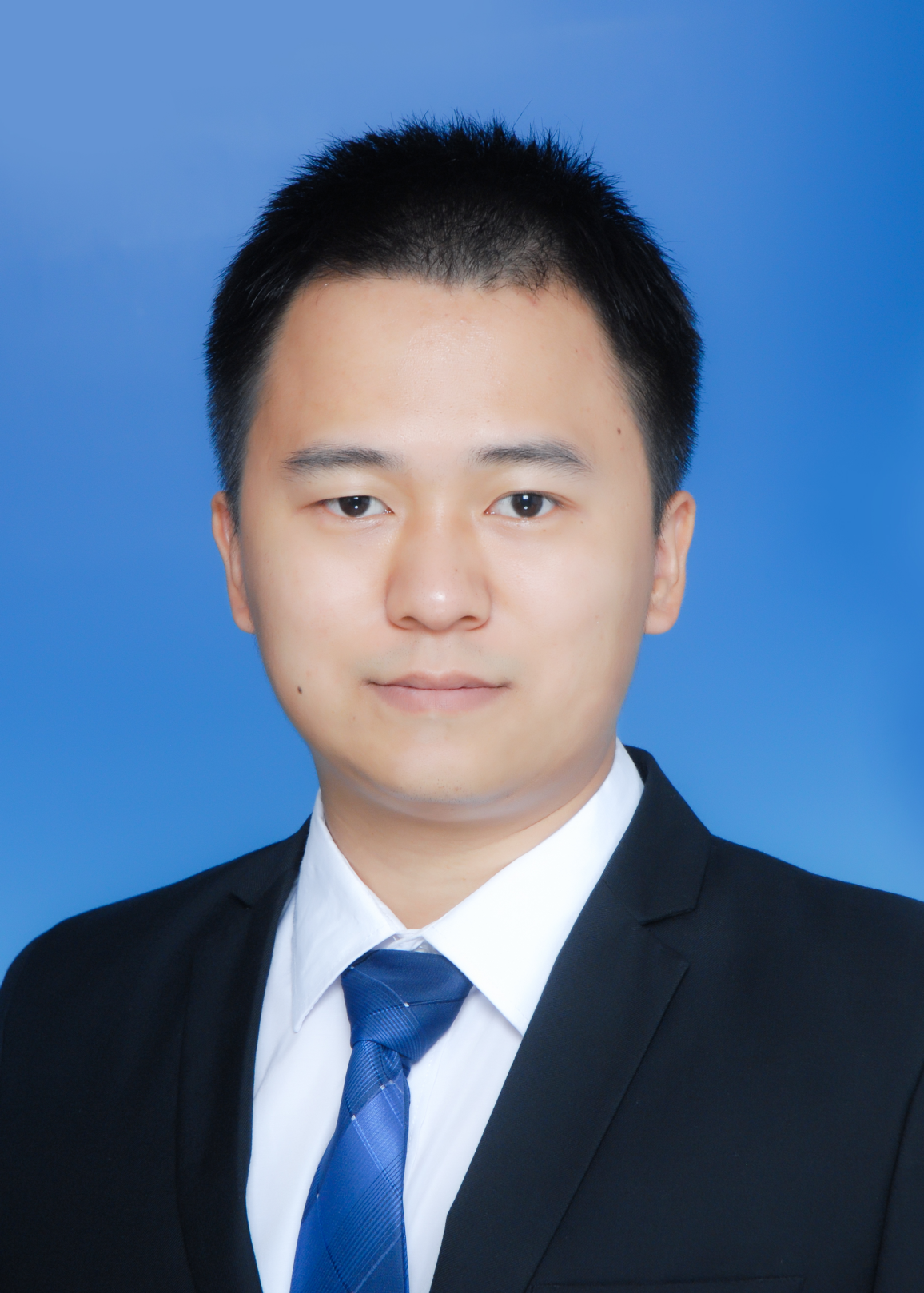}}]{Ruisi He }
 (S'11-M'13-SM'17) received the B.E. and Ph.D. degrees from Beijing Jiaotong University (BJTU), Beijing, China, in 2009 and 2015, respectively.

Dr. He is currently a Professor with the School of Electronics and Information Engineering, BJTU. Dr. He has been a Visiting Scholar in Georgia Institute of Technology, USA, University of Southern California, USA, and Universit\'e Catholique de Louvain, Belgium. His research interests include wireless propagation channels, 5G and 6G communications. He has authored/co-authored 8 books, 5 book chapters, more than 200 journal and conference papers, as well as several patents.

Dr. He has been an Editor of the IEEE Transactions on Communications, the IEEE Transactions on Wireless Communications, the IEEE Transactions on Antennas and Propagation, the IEEE Antennas and Propagation Magazine, the IEEE Communications Letters, the IEEE Open Journal of Vehicular Technology, and a Lead Guest Editor of the IEEE Journal on Selected Area in Communications and the IEEE Transactions on Antennas and Propagation. He served as the Early Career Representative (ECR) of Commission C, International Union of Radio Science (URSI). He received the URSI Issac Koga Gold Medal in 2021, the IEEE ComSoc Asia-Pacific Outstanding Young Researcher Award in 2019, the URSI Young Scientist Award in 2015, and several Best Paper Awards in IEEE journals and conferences.
\end{IEEEbiography} 
 
\begin{IEEEbiography}[{\includegraphics[width=1in,height=1.25in,clip,keepaspectratio]{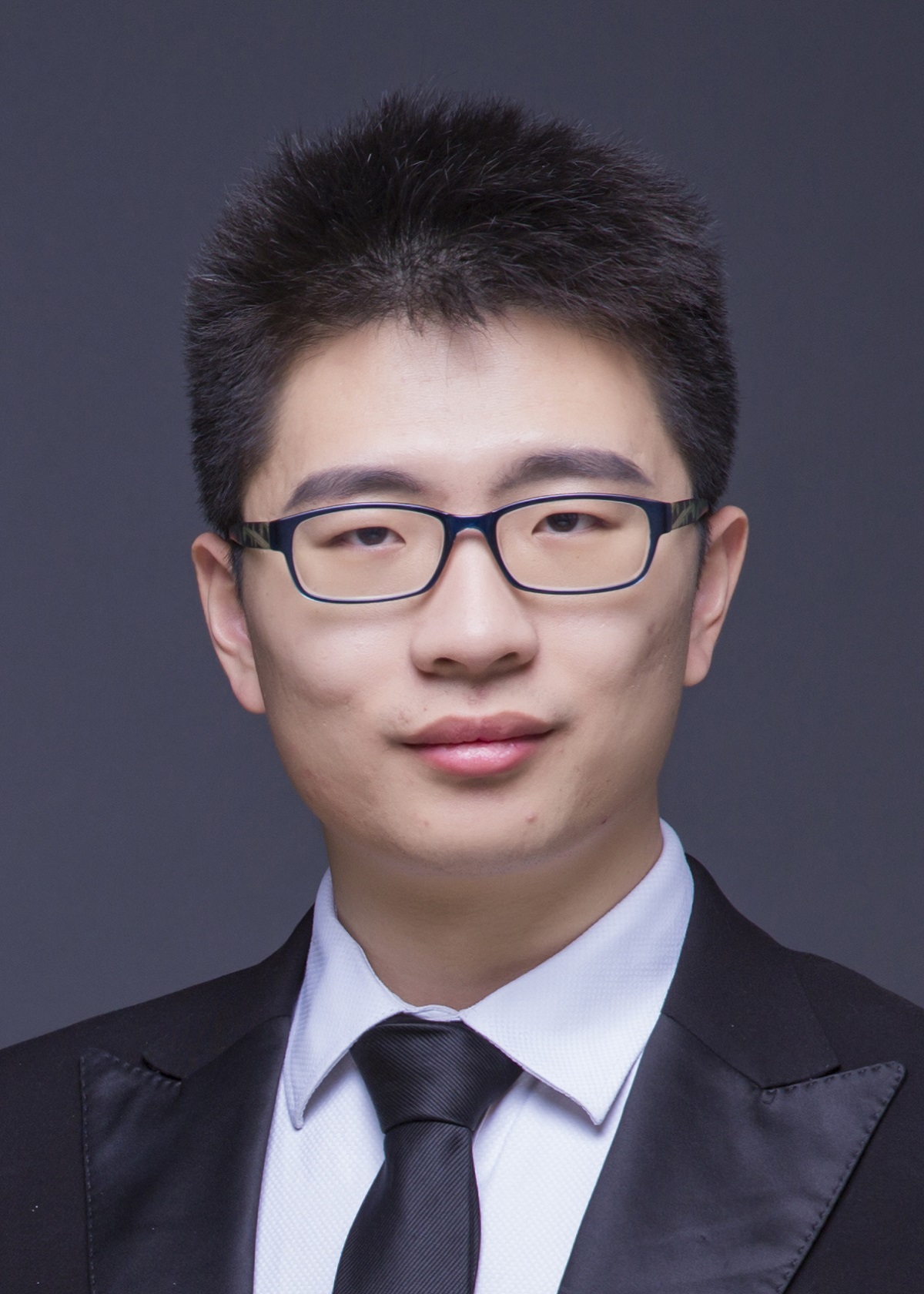}}]{Mi Yang }
 (S'17-M'21) received the M.S. and Ph.D. degrees from Beijing Jiaotong University, Beijing, China, in 2017 and 2021, respectively. He is currently an associate professor with the School of Electronic and Information Engineering, Beijing Jiaotong University. His research interests are focused on wireless channel measurement and modeling, vehicular and railway communications, and AI-enabled channel research.
\end{IEEEbiography}
 
\begin{IEEEbiography}[{\includegraphics[width=1in,height=1.25in,clip,keepaspectratio]{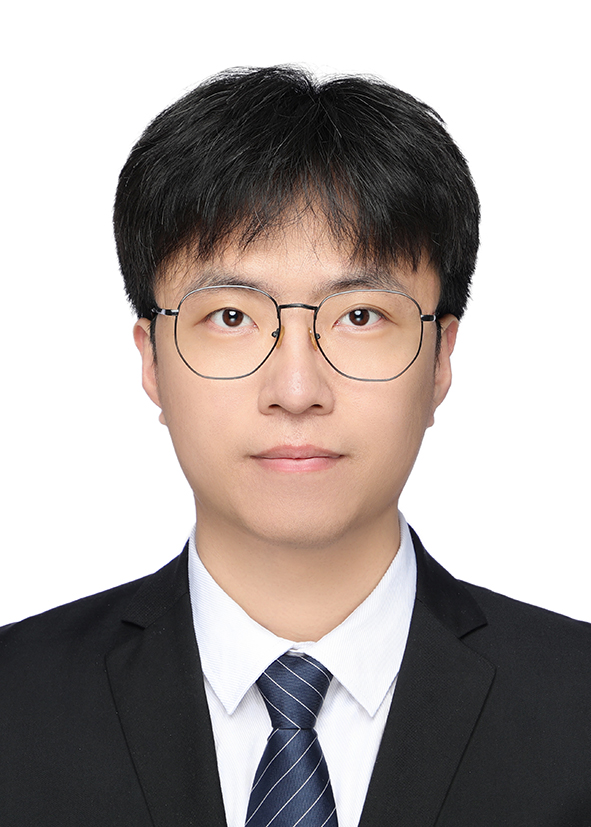}}]{Zhengyu Zhang } 
(S'21) received the M.S. and Ph.D degrees in electronic and communication engineerin from Beijing Jiaotong University, China, in 2021 and 2025, respectively, where he is currently pursuing the Ph.D. degree in information and communication systems. His research interests include ISAC, wireless communications, channel characterization and modeling, deep learning, etc.
\end{IEEEbiography} 
 
\begin{IEEEbiography}[{\includegraphics[width=1in,height=1.25in,clip,keepaspectratio]{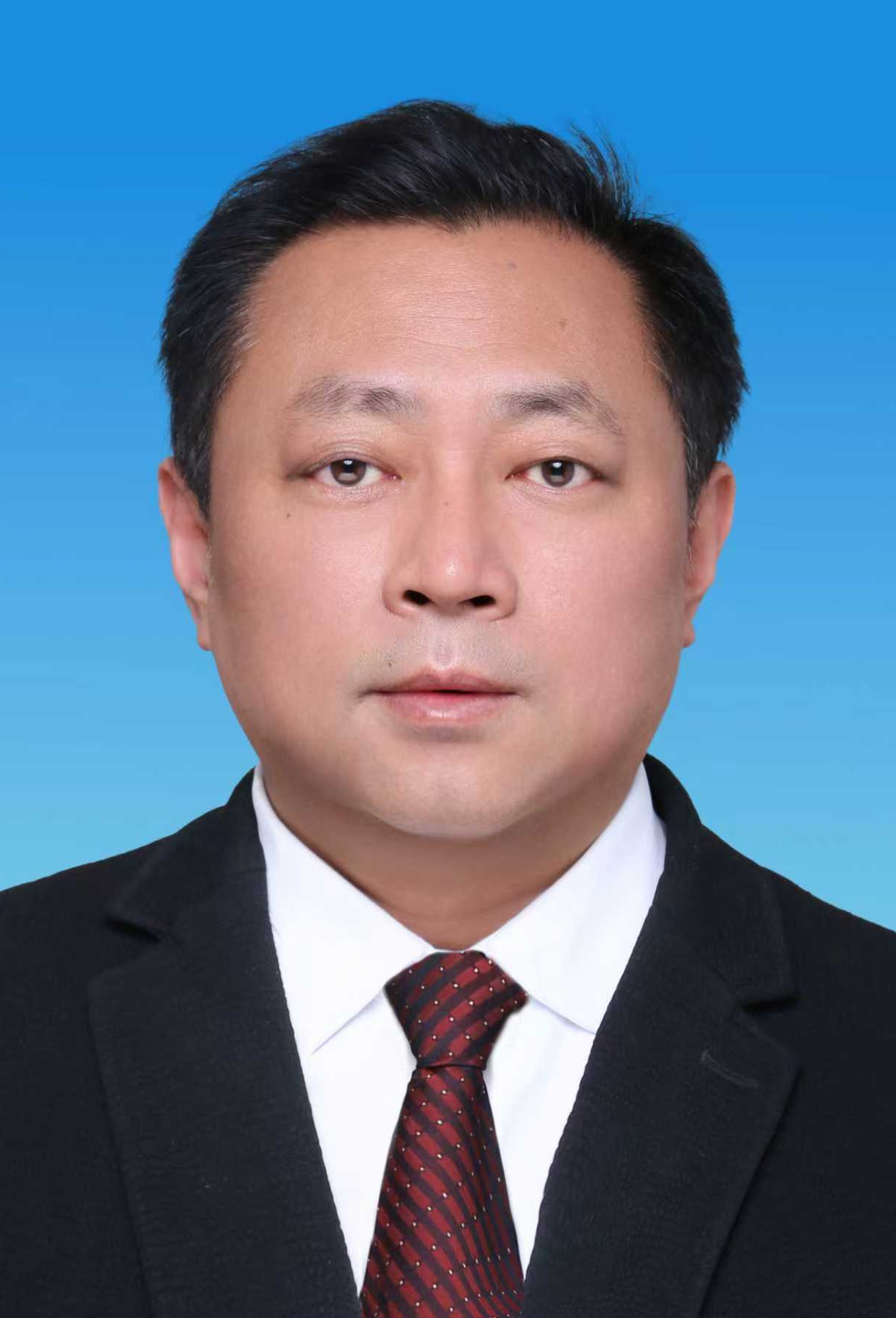}}]{Bo Ai }
(M'00-SM'10-F'21) received the M.S. and Ph.D. degrees from Xidian University, Xi’an, China, in 2002 and 2004, respectively. He was with Tsinghua University, Beijing, China, where he was an Excellent Postdoctoral Research Fellow in 2007. 
He is currently a Professor and an Advisor of Ph.D. candidates with Beijing Jiaotong University, Beijing. He has authored or coauthored six books and 270 scientific research papers, and holds 26 invention patents in his research areas. His interests include the research and applications of next generation wireless communication technology, radio propagation and channel modeling, and railway wireless communication systems.

He was as a Co-chair or a Session/Track Chair for many international conferences such as the 9th International Heavy Haul Conference (2009); the 2011 IEEE International Conference on Intelligent Rail Transportation; HSRCom2011; the 2012 IEEE International Symposium on Consumer Electronics; the 2013 International Conference on Wireless, Mobile and Multimedia; IEEE Green HetNet 2013; and the IEEE 78th Vehicular Technology Conference (2014). He has received many awards such as the Qiushi Outstanding Youth Award by HongKong Qiushi Foundation, the New Century Talents by the Chinese Ministry of Education, the Zhan Tianyou Railway Science and Technology Award by the Chinese Ministry of Railways, and the Science and Technology New Star by the Beijing Municipal Science and Technology Commission.
\end{IEEEbiography}

\begin{IEEEbiography}[{\includegraphics[width=1in,height=1.25in,clip,keepaspectratio]{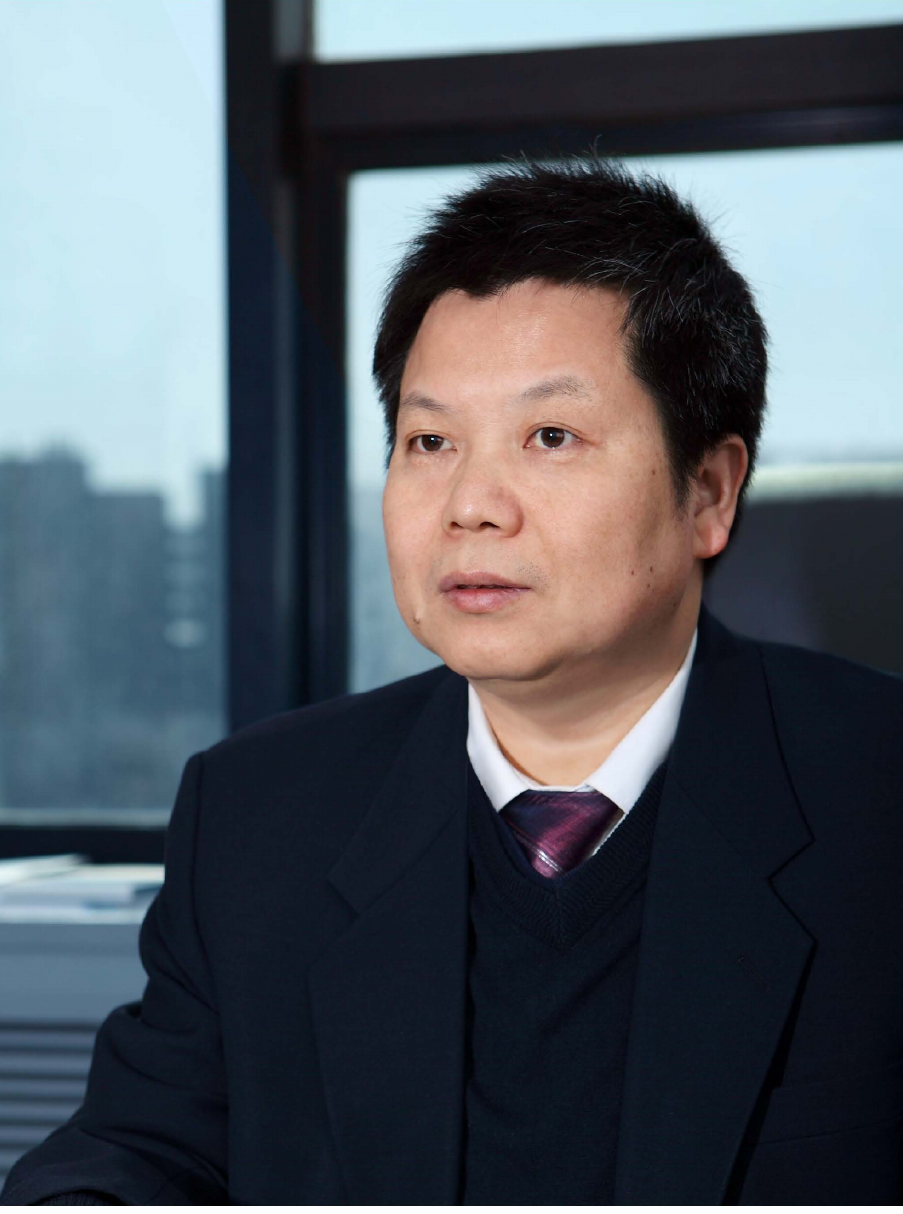}}]{Zhangdui Zhong } 
(SM'16-F'21) received the B.E. and M.S. degrees from Beijing Jiaotong University, Beijing, China, in 1983 and 1988, respectively. He is currently a Professor and an Advisor to Ph.D. candidates with Beijing Jiaotong University. He is also the Director of the Innovative Research Team of the Ministry of Education, Beijing. He has authored or co-authored seven books, five invention patents, and more than 200 scientific research papers in his research area. His research interests include wireless communications for railways, control theory and techniques for railways, and GSM-R systems. His research has been widely used in railway engineering, such as Qinghai-Xizang railway, DatongQinhuangdao Heavy Haul railway, and many high-speed railway lines in China. He is an Executive Council Member of the Radio Association of China, Beijing; and the Deputy Director of the Radio Association, Beijing. He received the Mao Yisheng Scientific Award of China, the Zhan Tianyou Railway Honorary Award of China, and the top ten Science/Technology Achievements Award of Chinese Universities.
\end{IEEEbiography}

\vfill\pagebreak

\end{document}